\documentclass[a4paper,11pt]{article}

\usepackage{jheppub}
\usepackage[T1]{fontenc}
\usepackage[utf8]{inputenc}
\usepackage[american]{babel}
\usepackage{csquotes}
\usepackage{subcaption}
\usepackage{float}

\usepackage{lipsum}
\usepackage{graphicx}
\usepackage{dcolumn}
\usepackage{tabularx}
\usepackage{bm} 
\usepackage{amssymb}
\usepackage{amsmath}
\usepackage{booktabs}
\usepackage{multirow}
\usepackage{makecell}
\usepackage[colorlinks=true]{hyperref}
\usepackage{lineno}

\title{\boldmath {Reconstruction of the event vertex in the PandaX-III experiment with convolution neural network}}

\author[a]{Tao Li,}
\author[c,1]{Yu Chen,}
\author[b,d,1]{Shaobo Wang,\note{Corresponding author.}}
\author[b]{Ke Han,}
\author[b]{Heng Lin,}
\author[b]{Kaixiang Ni,}
\author[a,c]{Wei Wang}
 
\affiliation[a]{Sino-French Institute of Nuclear Engineering and Technology, Sun Yat-sen University, Zhuhai 519082, China}
\affiliation[b]{INPAC; Shanghai Laboratory for Particle Physics and Cosmology;
Key Laboratory for Particle Astrophysics and Cosmology (MOE), \\
School of Physics and Astronomy, Shanghai Jiao Tong University, 800 Dongchuan Road, Shanghai {\rm 200240}, China} 
\affiliation[c]{School of Physics, Sun Yat-Sen University, 135 Xingang Xi Road, Guangzhou, {\rm 510275}, China}

\affiliation[d]{SPEIT~(SJTU-Paris Elite Institute of Technology), Shanghai Jiao Tong University, 800 Dongchuan Road, Shanghai, {\rm 200240}, China}

\emailAdd{litao73@mail.sysu.edu.cn}
\emailAdd{chenyu73@mail.sysu.edu.cn}
\emailAdd{shaobo.wang@sjtu.edu.cn}
\emailAdd{ke.han@sjtu.edu.cn}
\emailAdd{linheng@sjtu.edu.cn}
\emailAdd{1160274182@qq.com}
\emailAdd{wangw223@mail.sysu.edu.cn}

\date{\today}
\abstract{
The PandaX-III experiment uses a high-pressure xenon gaseous time projection chamber~(TPC) to search for the neutrinoless double beta decay~($0 \nu \beta \beta$) of $^{136}$Xe. 
The absence of the vertex position in the electron drift direction at which the event takes place in the detector limits the PandaX-III TPC's performance.
The charged particle tracks recorded by the TPC provide a possibility for vertex reconstruction. 
In this paper, a convolution neural network (CNN) model VGGZ0net is proposed for the reconstruction of vertex position. 
An 11~cm precision is achieved with the Monte Carlo simulation events uniformly distributed along a maximum drift distance of 120~cm. 
The electron loss during the drift under the different gas conditions is studied, and after the distance-based correction, the detector energy resolution is significantly improved.
The CNN model is also verified successfully using the experimental data of the PandaX-III prototype detector.
}

\begin{document}

\maketitle
\flushbottom

\section{Introduction}
\label{sec:CNN}
\begin{figure}[htp]
	\centering
	\includegraphics[width=.5\textwidth]{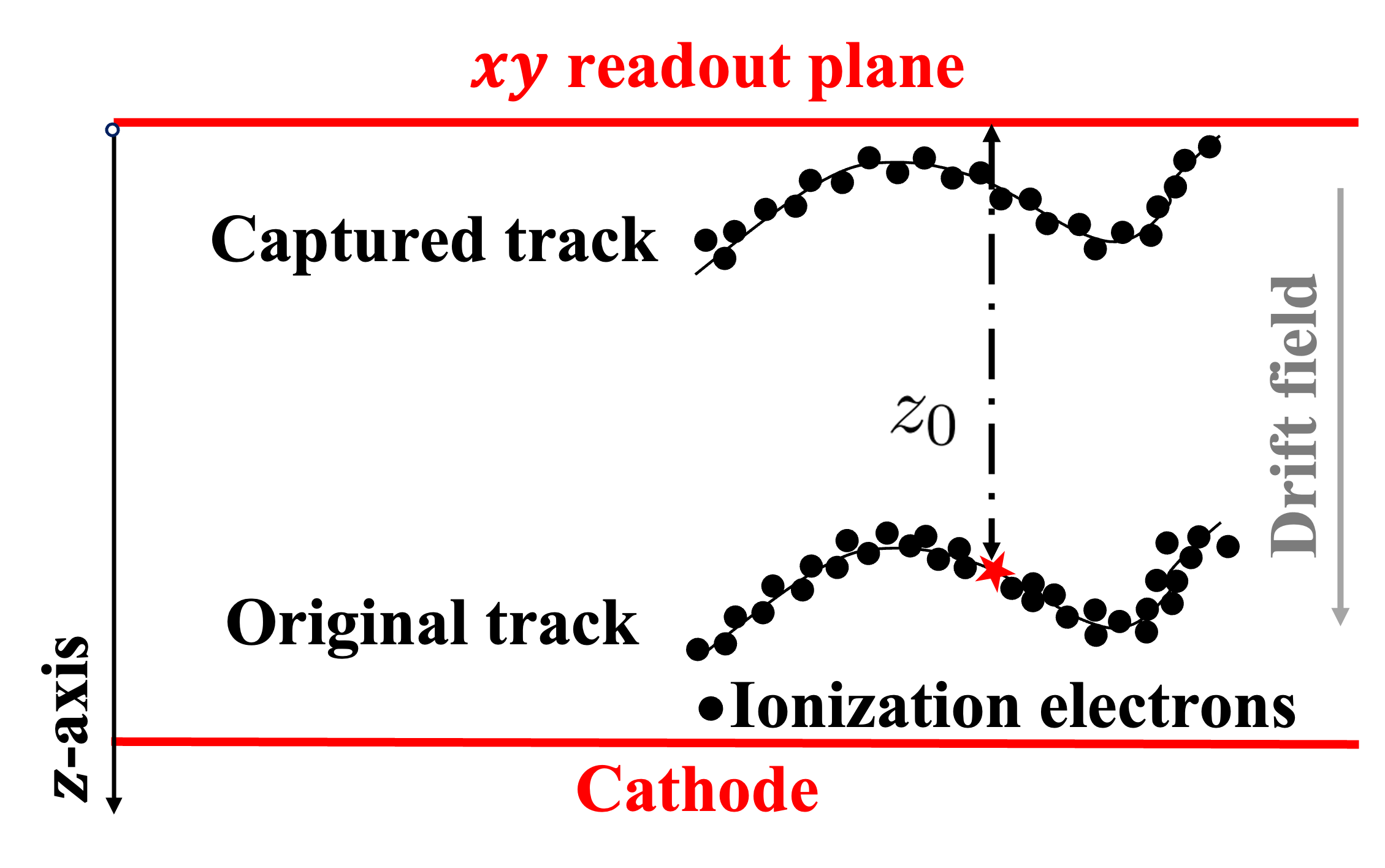}
	\caption{The working principle of the gaseous TPC. 
	An energetic charged particle gen- erates energy depositions by ionizing the working gas and leaves a three-dimensional track. The ionization electrons drift to the detector readout plane and are collected. An example of the original track and the captured track shows the electron loss during the drift process. The star on the original track represents the vertex $z_0$.}
	\label{fig:TPCconstruction}
\end{figure}

Neutrinoless double beta decay~($0\nu\beta\beta$) processes are sensitive probes of physics beyond the standard model of particle physics.
A discovery of such a process in experiments would directly indicate the Majorana nature of neutrinos and provide a mechanism for the violation of lepton number conservation~\cite{Avignone:2007fu}.
Many collaborations are searching for the $0\nu\beta\beta$ with different isotopes, 
such as $^{76}$Ge~\cite{GERDA:2020xhi}, $^{130}$Te~\cite{CUORE:2022jto}, and $^{136}$Xe~\cite{KamLAND-Zen:2022tow, EXO-200:2019rkq} through analyzing the energy spectrum around the Q-value~($Q_{\beta \beta}$), 
the total energy released by the decay and carried by the two emitted electrons. 
In recent years, the gaseous Time Projection Chambers~(TPCs) have emerged as powerful tools for the $0\nu\beta\beta$ searches~\cite{Gomez-Cadenas:2019ges}.
Figure~\ref{fig:TPCconstruction} shows the working principle of TPCs.
When a charged particle propagates through the medium gas, it would deposit energy along its trajectory by ionizing the gas atoms. 
Primary ionization electrons drift along the electric field in the vertical direction, and are collected at the $xy$ readout plane.
The drift time of ionization electrons will provide $z$ position measurement. 
The vertex position in the drift direction at which the event takes place is defined as $z_0$.
For example, the $z_0$ of a $0\nu\beta\beta$ event is the location of decay occurrence.
Both the energy deposition and the particle trajectory could be recorded by TPC, which is expected to reveal a comprehensive detector response and enhance the sensitivity of $0\nu\beta\beta$ detection through the topological analysis~\cite{NEXT:2019gtz,Galan:2019ake,Li:2021viv}.
However, a portion of the ionization electrons will be reabsorbed due to the gas impurities during the drifting, resulting in the so-called “electron attachment effect” characterized by the electron lifetime~\cite{Thomas:1987zz}. 
The electron loss will worse the detector energy resolution $\Delta E$ and limit the projection sensitivity of $0\nu\beta\beta$ search $S_{T_{1/2}}^{0 \nu \beta \beta} \propto \sqrt{\frac{1}{\Delta E}}$.
Only if both the electron lifetime and $z_0$ are known, the energy resolution could be corrected event by event.
Therefore, the electron lifetime and the $z_0$ reconstruction are the critical parameters for searching $0 \nu \beta \beta$ with TPC.

The PandaX-III collaboration~\cite{Chen:2016qcd} has proposed the construction of a high-pressure gaseous xenon TPC to search for $0\nu\beta\beta$ of $^{136}$Xe. 
Xenon is mixed with 1\% trimethylamine~(TMA) to improve signal quality of the TPC.
The use of TMA provides a stable operation of the Micromegas~\cite{NEXT:2015exs} readout plane of detector at high pressure, as well as a reduction of the electron diffusion and a quenching of the primary xenon scintillation light while simultaneously displaying Penning effect~\cite{Herrera:2014fsb}. 
However, $z_0$ is not directly available due to the loss of scintillation light signal in such a xenon-TMA mixture.
In this work, we focus on the $z_0$ reconstruction using a method of the Convolution Neural Network~(CNN), 
which has been widely used in particle physics experiments for signal recognition and physical feature extraction~\cite{Aurisano:2016jvx, MicroBooNE:2020hho, Qiao:2018edn, NEXT:2020jmz}. 
In the PandaX-III TPC, as the ionization electrons drift toward the anode along the drift field, 
the diffusion effect carries the $z_0$ information of events: the longer drift distance a specific particle goes through, the more diffused its track shows. 
Therefore, the charge and position dispersion of  ionization electrons on a track is perceived by the CNN to realize the $z_0$ reconstruction.

The paper is organized as follows. The geometry of the PandaX-III detector as well as the correlation between $z_0$ and the electron diffusion effect is presented in Section~\ref{sec:T0}. 
The realization of the CNN model and the evaluation of $z_0$ reconstruction through the simulated events are described in Section~\ref{sec:CNN}.
In Section~\ref{sec:Kr}, we verified the effectiveness of the model using the experimental data of the PandaX-III prototype detector~\cite{Lin:2018mpd}. 
Finally, a brief summary is given in Section~\ref{sec:conclusion}. 

\section{The PandaX-III detector and event vertex}
\label{sec:T0}

\begin{figure}[htp]
	\centering
	\includegraphics[width=.55\textwidth]{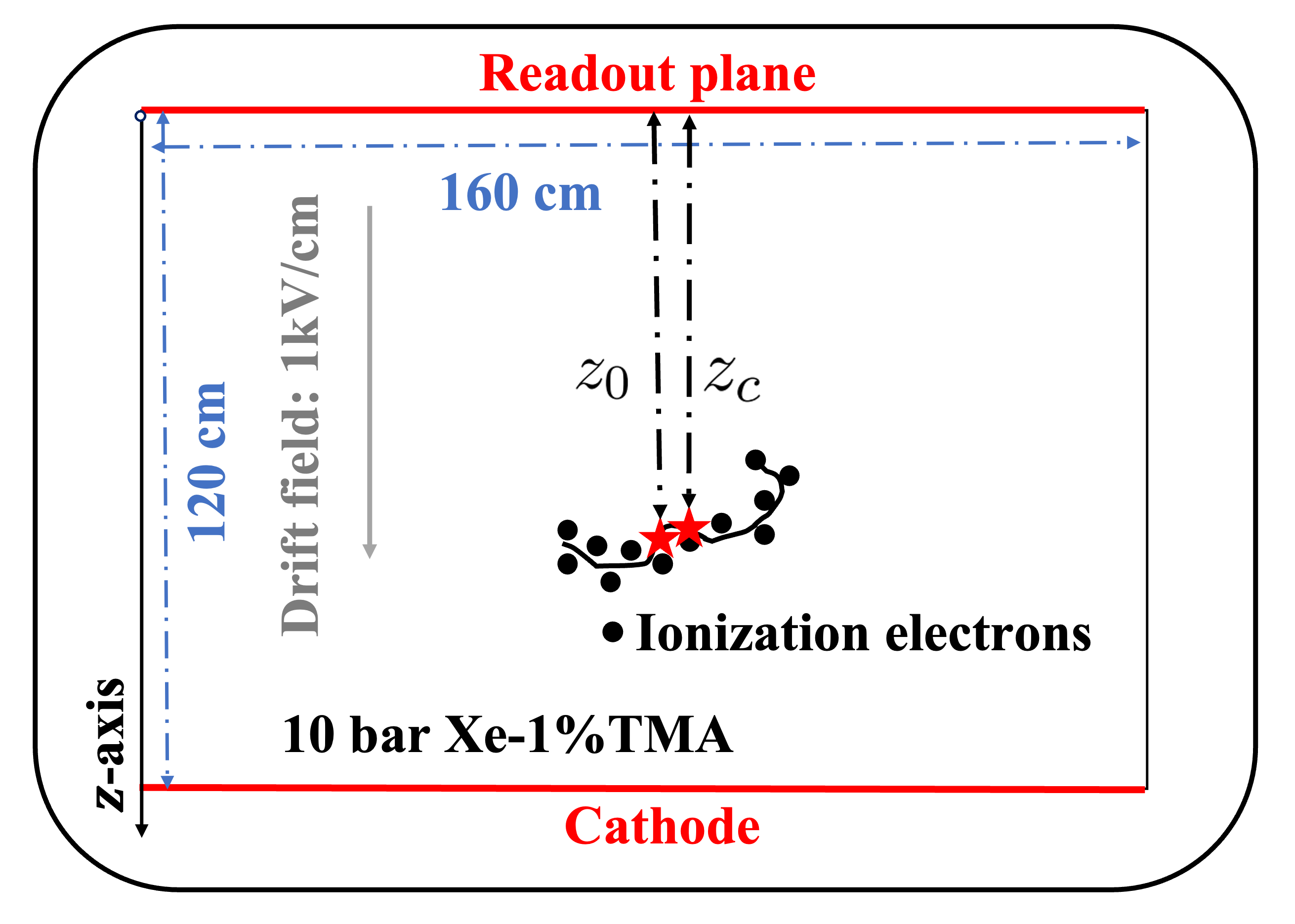}
	\caption{Illustration of the PandaX-III TPC geometry. The TPC has a cylindrical active volume with a height of 120~cm and a diameter of 160~cm.
	An example $0\nu\beta\beta$ event track is shown. The $z_0$ is the vertex position at which the decay takes place. 
	The $z_c$ is the position of the charge center, which is chosen to substitute $z_0$ as the reconstruction target.}
	\label{fig:TPC}
\end{figure}

The specific design of the PandaX-III detector is presented in Ref.~\cite{Wang:2020owr}.
As shown in Figure~\ref{fig:TPC}, the TPC will be a cylindrical active volume with a height of 120~cm and a diameter of 160~cm, 
containing 140~kg enriched $^{136}$Xe mixed with 1\% TMA at 10~bar operating pressure. 
The charge readout plane consists of a tessellation of fifty-two 20$\times$20~cm$^2$ Micromegas modules with a series of 3~mm pitch strips readout in the horizontal direction. 
The expected energy resolution of the PandaX-III detector is about 3\% Full Width at Half Maximum~(FWHM) at the $Q_{\beta \beta}$ (2.458~MeV).
In the PandaX-III TPC, the vertex position of event $z_0$ cannot be measured directly, which is a crucial issue of the energy resolution considering a drift distance of 120~cm. 
A precise $z_0$ reconstruction is necessary for a high-pressure TPC aiming for the energy resolution of zero-attachment.
In this work, we use the position of event charge center of track $z_c$ to replace $z_0$. Once $z_c$ is reconstructed, the absolute $z$ position of all the hit points in the track becomes available.

\begin{figure}[h]
	\centering
	\includegraphics[width=1.\textwidth]{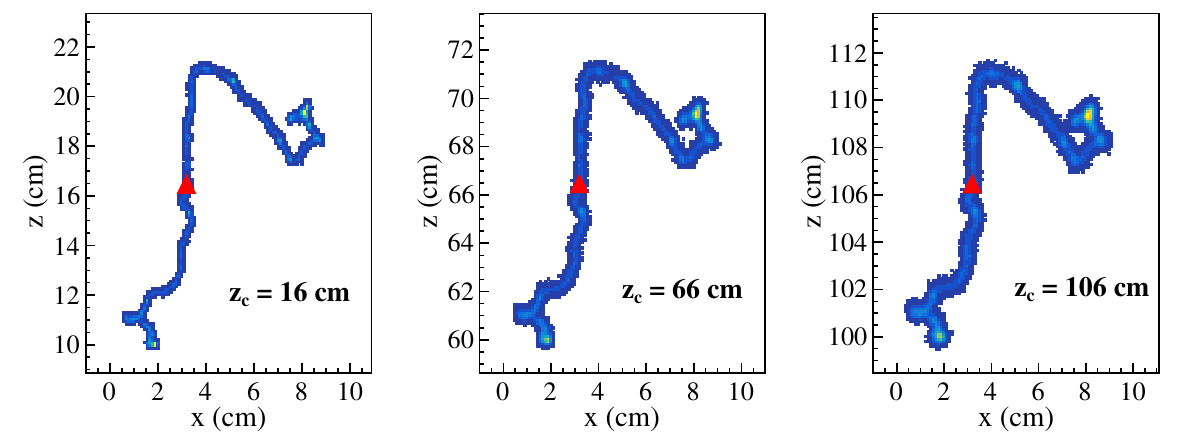}
	\caption{The simulated tracks (projection in $xz$ plane) for the same $0\nu \beta \beta$ event with different $z_c$ (shown in red triangles) of 16~cm, 66~cm, and 106~cm.
	}
	\label{fig:diffusion tracks}
\end{figure}

While drifting to the readout plane, ionization electrons diffuse transversely and longitudinally, leading to
the blurring of the tracks. In the PandaX-III TPC, the drift velocity of ionization electron $v_{drift}$ is 1.86~mm/$\mu$s by applying a drift field of 1~kV/cm, 
and the transverse (longitudinal) diffusion coefficient is 1.0~(1.5)$\times$10$^{-2}$~cm$^{1/2}$~\cite{Li:2021viv}. The diffusion could degrade the energy and track reconstruction. 
However, the degree of such dispersion could reveal its vertex position.
As shown in Figure~\ref{fig:diffusion tracks}, the same event with different $z_c$ within the detector is simulated, a coarser track could be observed after a longer drift distance. 
Hence, we propose a CNN-based method to reconstruct $z_c$ and estimate electron lifetime by extracting the track features.

\section{CNN and network architectures}
\label{sec:CNN}

A CNN regression model, named \emph{VGGZ0net}, is built to predict the $z_c$ of events.
The VGGZ0net is based on the VGG16 classification model proposed by the Visual Geometry Group at Oxford University in 2014~\cite{simonyan2014very}.
In the PandaX-III, VGGZ0net is designed to solve a regression problem considering the continuous distribution of event location along the drift distance of 120~cm. 

\subsection{Simulation and data production}
\label{sec:simulation}
\begin{figure}[h]
    \begin{subfigure}[t]{.60\textwidth}
        \centering
        \includegraphics[width=.95\textwidth]{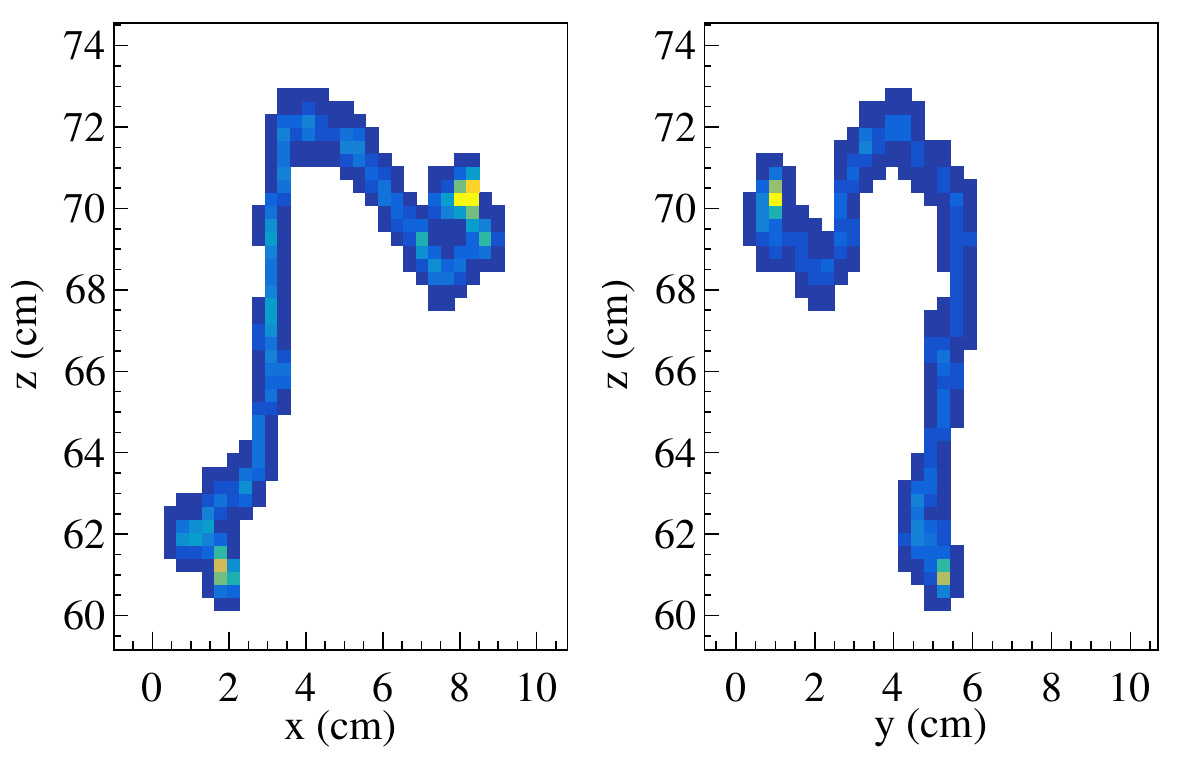}
        \caption{}
        \label{fig:0vbb_readout}
    \end{subfigure}
    \begin{subfigure}[t]{.35\textwidth}
        \centering
        \includegraphics[width=1.\textwidth]{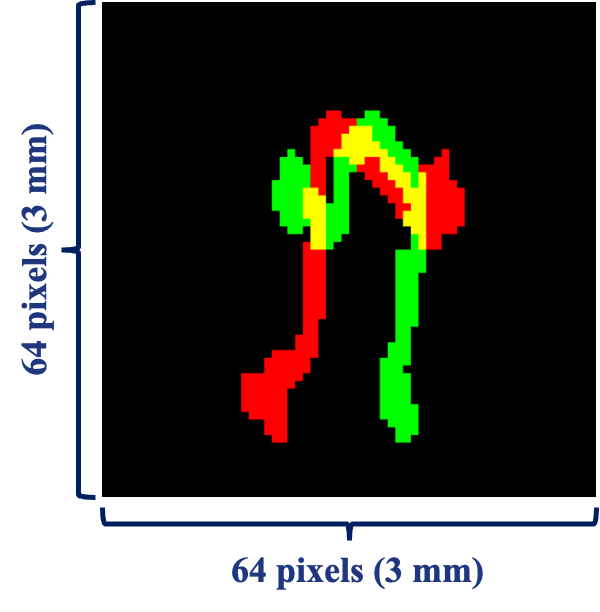}
        \caption{}
        \label{fig:0vbb_h5Png_hit}
    \end{subfigure}
    \caption{Demonstration of the dataset production for the simulated $0\nu\beta\beta$ event: 
        (a) two-dimensional tracks on $xz$~(Left) and $yz$~(Right) planes;
        (b) the converted RGB image as the input of CNN:  all the triggered pixels shown here are set to the maximum values for better visualization.} 
    \label{fig:0vbb_png}
\end{figure}

Monte Carlo~(MC) simulation datasets are generated using the REST framework~\cite{Altenmuller:2021slh} to optimize and validate the model.
The $0\nu\beta\beta$ events of $^{136}$Xe, together with a corresponding major background, the $\gamma$ events of 2.615~MeV from $^{232}$Th decay chain, are simulated.
The events are firstly produced by $restG4$ of REST, based on the Geant4 framework~\cite{Agostinelli:2002hh}. 
Then the detector responses, including electron diffusion, electron attachment, energy smearing, and Micromegas readout plane scheme, are simulated by $restManager$.
More details of the simulation can be found in Ref.~\cite{Li:2021viv}.
The simulation is further optimized according to the actual detector and electronics response, including a sampling rate of 5~MHz, a waveform shaping time of 5~$\mu$s, and an electronic noise level of 0.3~fC.
The triggered channels provide the $x$ and $y$ positions, and the sample points of each waveform provide a relative $z$ position once drift velocity and relative drift time are given.
Figure~\ref{fig:0vbb_readout} shows the tracks of such an event on the $xz$ and $yz$ planes, respectively. 
Afterwards, the energy of tracks on the $xz$ and $yz$ plane is normalized, and the tracks are filled into red ($xz$) and green ($yz$) channels of the input RGB image to VGGZ0net, as shown in Figure~\ref{fig:0vbb_h5Png_hit}. 
The image format is 64$\times$64 pixels with a size of 3$\times$3~mm$^2$ each. 
The image size is 19.2~cm in each direction can ensure the whole track of $0\nu\beta\beta$ being inside the image area. 

\begin{figure}[h]
    \centering
    \includegraphics[width=.6\textwidth]{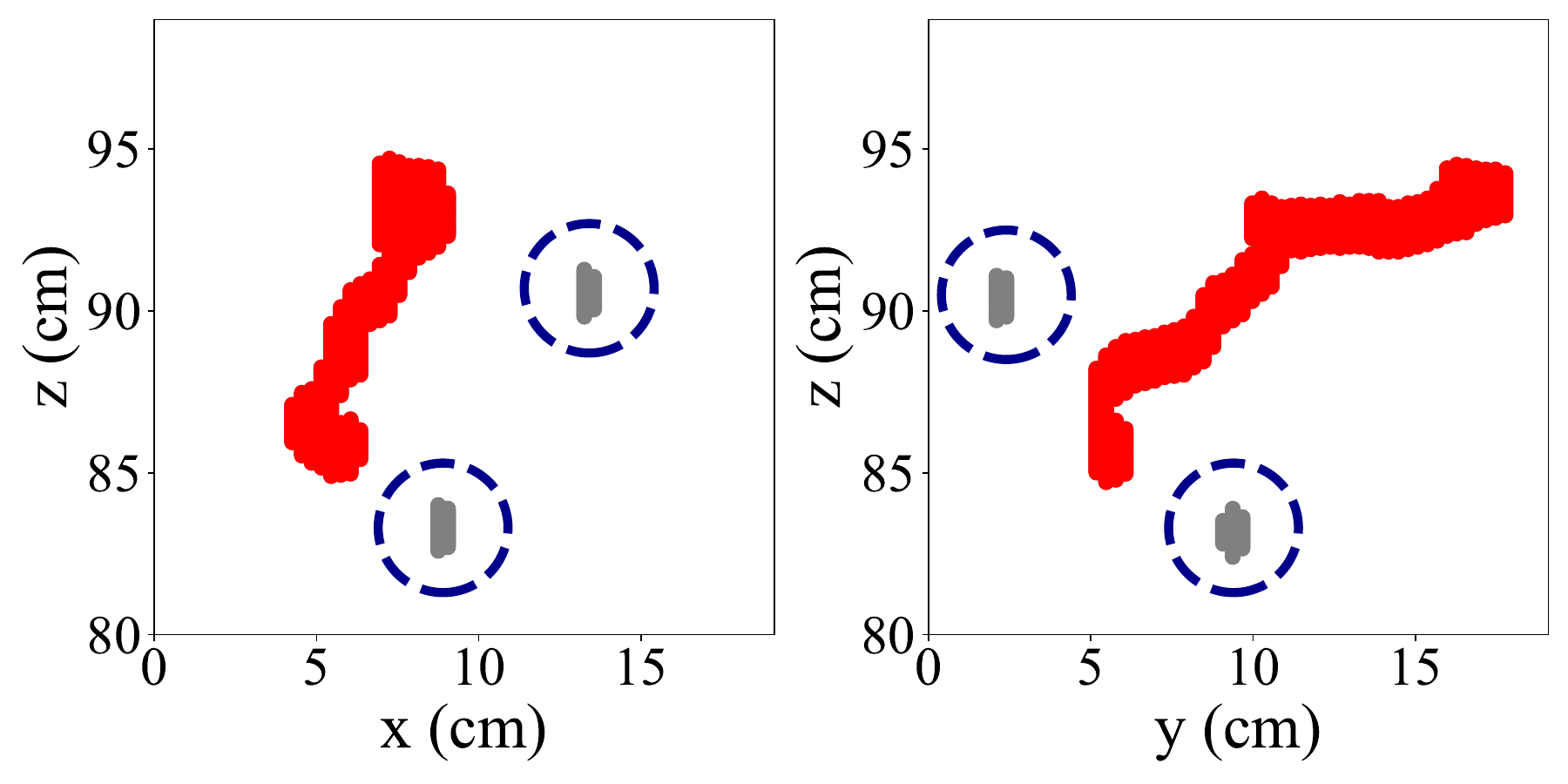}
    \caption{An example of simulated $\gamma$ event of $^{232}$Th. The tracks on the $xz$~(Left) and $yz$~(Right) planes are shown. The subordinate tracks have been marked with blue dotted circles. The DBSCAN algorithm~\cite{sander1998density} extracts the principle track in red.}
    \label{fig:Th232_DBSCAN}
\end{figure}

The events with the total energy of more than 2.0 MeV, which is the important energy range for $0\nu\beta\beta$ search, are selected as the dataset. 
The principle tracks of the dataset, which are featuring a continuous track in each event with significant energy deposition, are labeled by $z_c$.
In the $0\nu\beta\beta$ and $\gamma$ events, several subordinate tracks may be generated, especially in the $\gamma$ events where Compton scattering could generate multiple tracks.
To simplify the data processing, the principle track within each event is selected and its $z_c$ will be predicted by VGGZ0net. Once $z_c$ is derived, all the subordinate tracks can be located through their relevant distance. 
The DBSCAN clustering algorithm \cite{sander1998density} is used to pick out the principle track as a preprocessing step. An example of the $\gamma$ dataset is shown in Figure~\ref{fig:Th232_DBSCAN}.
We produced the scenarios with different electron lifetimes to study $z_c$ reconstruction and energy correction.
For the ideal scenario of zero-attachment, the energy spectrum is smeared by a Gaussian function which represents the energy resolution of 3.3\% FWHM at $Q_{\beta \beta}$ of $^{136}$Xe. 
Electron lifetime is endowed with introducing a coefficient of electron adsorption in gas based on the zero-attachment scenario.

\subsection{Network architectures and performance}
\label{sec:network}
\begin{figure}[h]
    \centering
    \includegraphics[width=1.\textwidth]{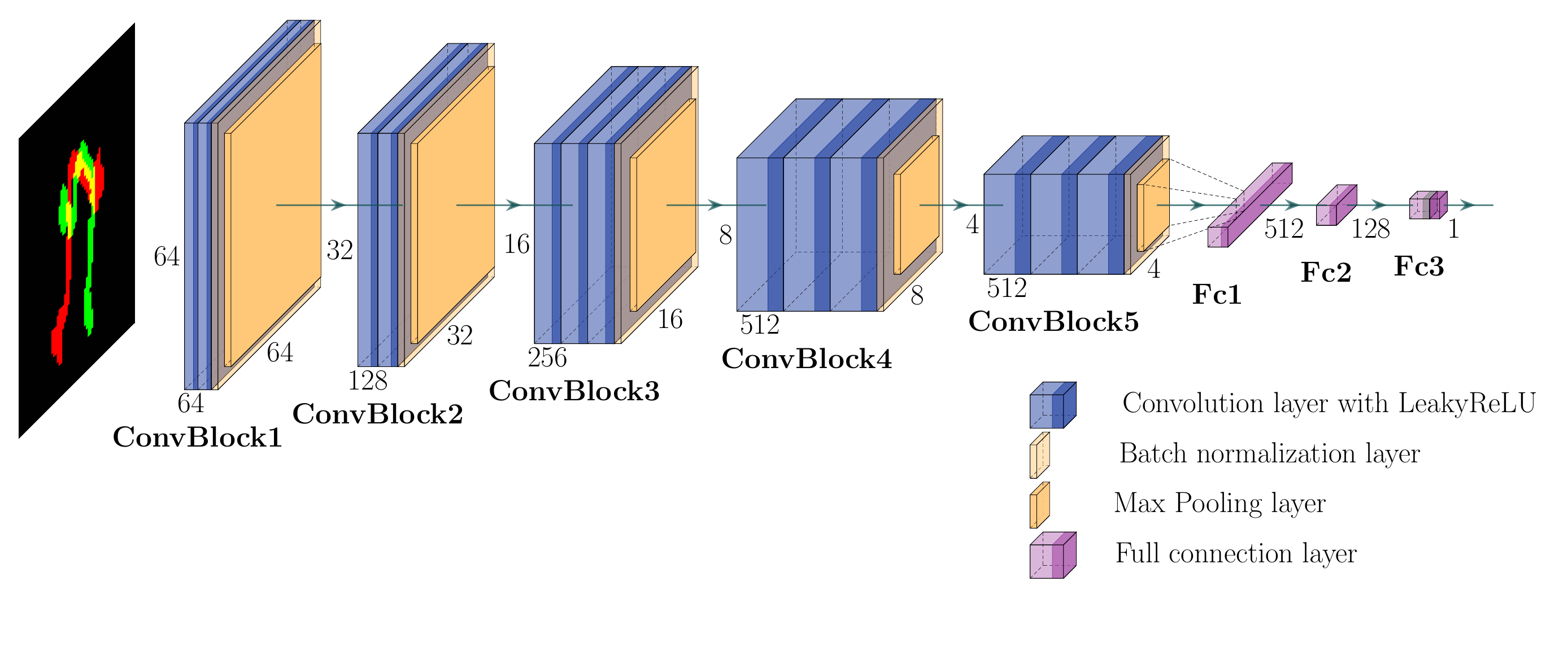}
    \caption{The structure of VGGZ0net.}
    \label{fig:model_structure}
\end{figure}

The structure of VGGZ0net is represented in Figure~\ref{fig:model_structure}, consisting of 5 sets of convolution blocks as the feature extraction part. 
The size of each layer is shown in Figure~\ref{fig:model_structure}. LeakyReLU~\cite{maas2013rectifier} is chosen as the activation function of the convolution layers.
Thereinto, batch normalization layer is utilized to improve accuracy and speed up training~\cite{bjorck2018understanding}.
After the feature extraction part, a fully connected part is designed with 2 fully connected layers with LeakyReLU as the activation function, as well as a final connected layer and a linear activation function. 
A mean square error~(MSE) loss function is adopted for the regression requirements. A stochastic gradient descent method, Adam~\cite{kingma2014adam}, is chosen as the optimizer.

\begin{table}[h]
    \centering
    \renewcommand\arraystretch{1.}
    \caption{The tuning results of the optimal hyper-parameters in VGGZ0net.}
    \begin{tabular}{*{2}{c}}
        \toprule
        Hyper-parameters & tuning results \\
        \midrule
        Label range of dataset & (-1, 1) \\
        Batch size per epoch & 128 \\
        Negative slope coefficient of leakyReLU & 0.6 \\
        Learning rate of Adam optimizer & 7$\times$10$^{-5}$ \\
        \bottomrule
    \end{tabular}
    \label{tab:optimal_hp}
\end{table}

The model is first trained through a supervised learning process based on the MC $0\nu\beta\beta$ events of the zero-attachment scenario. 
Both the training and validation datasets consist of 120,000 events.
A grid search method was conducted within four main hyper-parameters in VGGZ0net to optimize the model performance. The optimized values of hyper-parameters used are shown in Table~\ref{tab:optimal_hp}.

\begin{figure}[h]
    \centering
    \includegraphics[width=.75\textwidth]{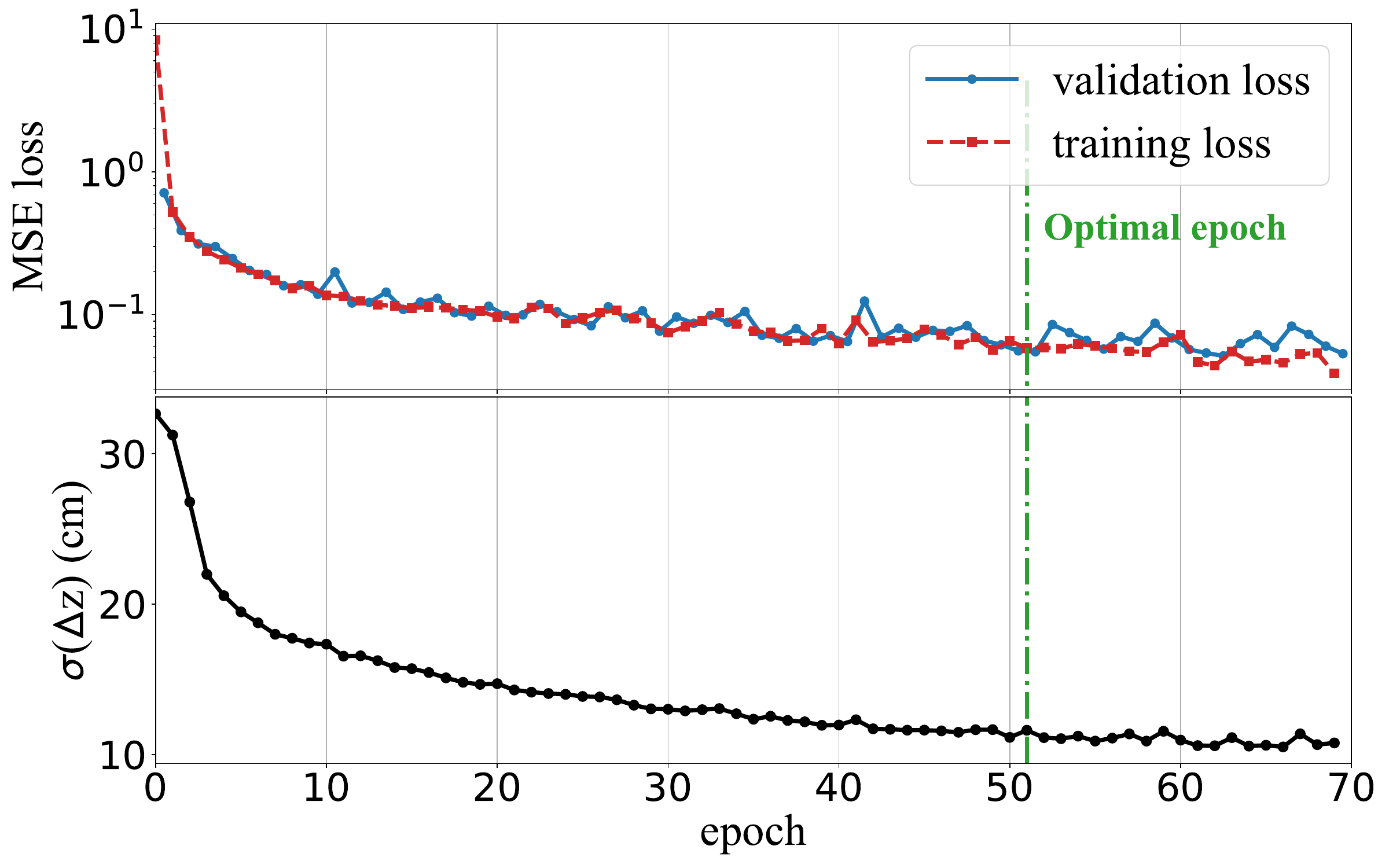}
    \caption{The evolution of the MSE loss function (Top).  The $\sigma(\Delta z)$ evolution on the validation dataset (Bottom). 
    The values of hyper-parameters in Table~\ref{tab:optimal_hp} are used, and the optimal model at around $50^{th}$ epoch is chosen.}
    \label{fig:0vbb_model_process} 
\end{figure}
The results of VGGZ0net on $0 \nu \beta \beta$ training and validation datasets are shown in Figure~\ref{fig:0vbb_model_process}.
We defined the regression error:
\begin{equation}
    \Delta z = \hat{z}_{c} - z_{c},
\end{equation}
where $ z_{c}$ is from the label of events, and $\hat{z}_{c}$ is that reconstructed by VGGZ0net.
The $\sigma(\Delta z)$ is the standard deviation of $\Delta z$. 
As shown in Figure~\ref{fig:0vbb_model_process}, both the training and the validation MSE loss of VGGZ0net are continuously reduced during the training process, resulting in $\sigma(\Delta z)$ decreasing gradually.
The optimal model is chosen conservatively around the $50_{th}$ epoch without overfitting. 

\begin{figure}[h]
    \begin{subfigure}[h]{1.\textwidth}
        \centering
        \includegraphics[width=.7\textwidth]{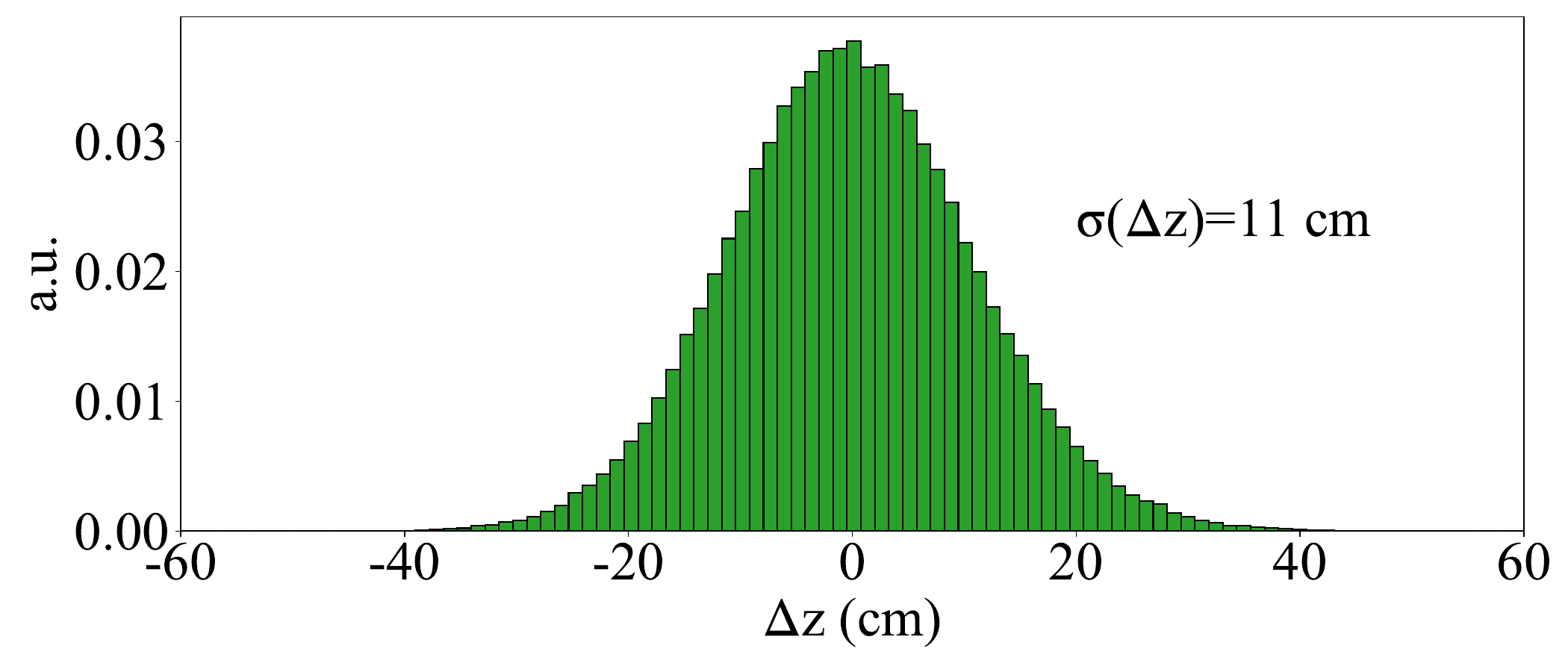}
    \end{subfigure}
    \begin{subfigure}[h]{1.\textwidth}
        \centering
        \includegraphics[width=.8\textwidth]{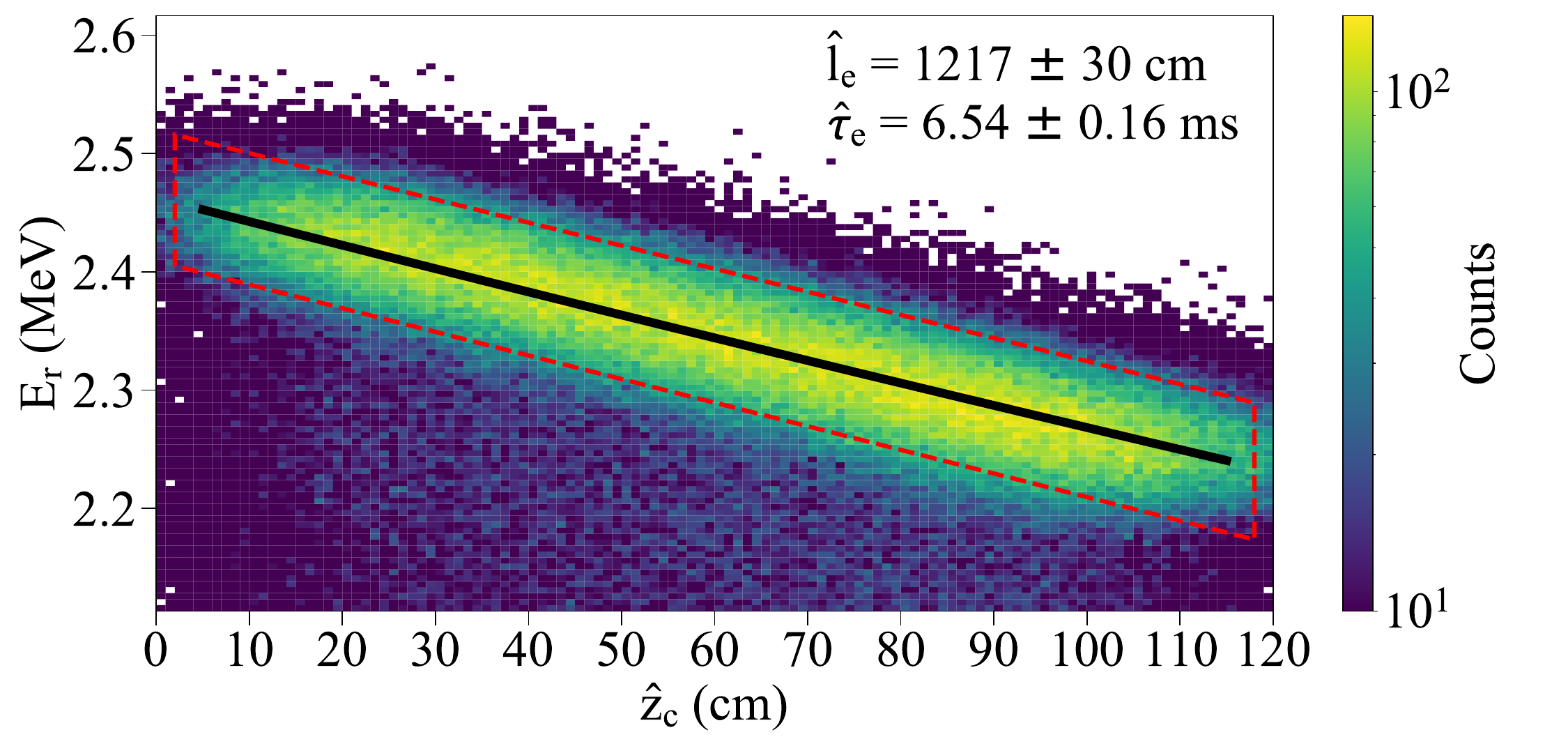}
    \end{subfigure}
    \caption{The prediction performance of VGGZ0net model in a dataset of $0\nu\beta\beta$ with $l_e$ of 1200~cm. The $z_c$ reconstruction precision $\sigma(\Delta z)$ is 11~cm (Top). 
    The spectrum of raw energy $E_r$ over $\hat{z}_c$ shows the electron loss along the drift distance (Bottom).
    The $\hat{l}_e$ of 1217$\pm$30~cm is obtained through fitting the energy band of  $Q_{\beta \beta}$ within the red dotted box region.
    The corresponding $\hat{\tau}_e$ is 6.54$\pm$0.16~ms.
    }
    \label{fig:0vbb_1200_val}
\end{figure}
The trained model is applied to the scenarios of $0\nu\beta\beta$ with the defined electron absorption distance $l_e$ of 1200~cm, which is equivalent to the electron lifetime $\tau_e$ of 6.45~ms ($\tau_e = l_e/v_{drift}$).
The $\sigma(\Delta z)$ obtained by VGGZ0net is about 11~cm for the events uniformly distributed within 120~cm distance, as shown in Figure~\ref{fig:0vbb_1200_val}~(Top).
The event energy distribution is plotted over the reconstructed $\hat{z}_{c}$ shown in Figure \ref{fig:0vbb_1200_val}~(Bottom). An exponential function is introduced to obtain the electron absorption distance and electron lifetime:
\begin{equation}
    E_{Q} = A_0 \cdot e^{- \hat{l}_e/\hat{z}_{c}} = A_0 \cdot e^{- \hat{\tau}_e/\hat{t}_{c}},
   \label{eq:decay}
\end{equation}
where the $E_{Q}$ is the fitted mean of detected $Q_{\beta \beta}$ peak in the TPC, and $A_0$ is an amplitude coefficient.
The fitted electron absorption distance $\hat{l}_e$ of 1217$\pm$30~cm conforms to the defined $l_e$ within one standard deviation. 
The $\hat{t}_{c}$ is the drift time related to $\hat{z}_{c}$, the corresponding electron lifetime $\hat{\tau}_e$ is 6.54 $\pm$ 0.16~ms.
Given the charge normalized images, our model is insensitive to the absolute energy of the event. 
Therefore, the trained model is also applicable to the dataset of $\gamma$; the $\sigma(\Delta z)$ by the same trained model is still about 11~cm. 
The reconstructed $\hat{l}_e$ is 1194$\pm$27~cm for the dataset of $\gamma$ with $l_e$ of 1200~cm.
Thus, it demonstrates the effectiveness and applicability of this method for the both $0\nu\beta\beta$ and $\gamma$ events in the same energy range.

\subsection{Energy correction}

\begin{figure}[h]
    \centering
    \includegraphics[width=.8\textwidth]{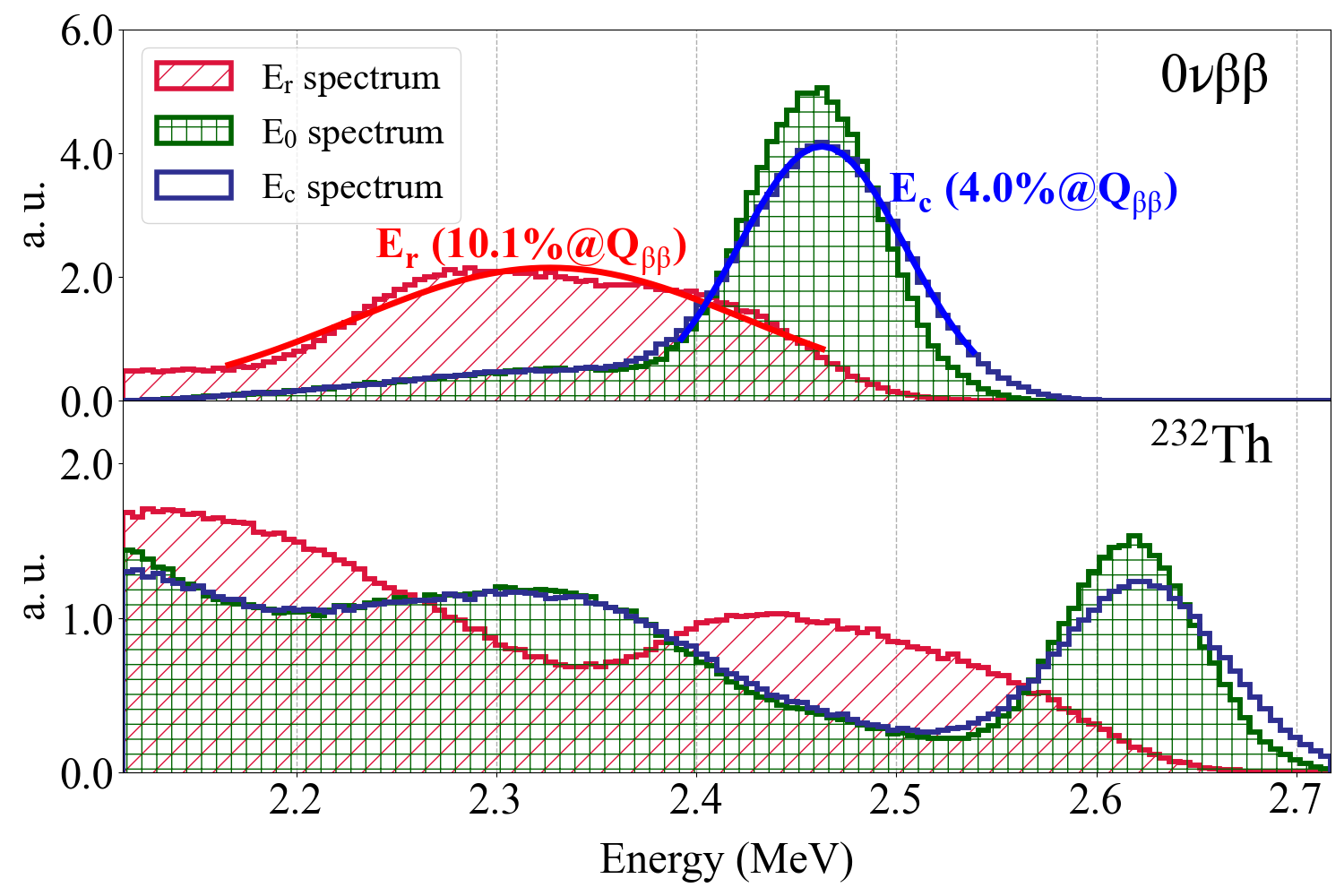}
    \caption{The energy spectra of simulated $0 \nu \beta \beta$ (Top)  and $\gamma$ from $^{232}$Th (Bottom) datasets. The $E_0$ represents the energy of the zero-attachment scenario, $E_r$ is the raw energy affected by the introduced electron attachment, 
    and $E_c$ is the energy corrected by VGGZ0net.
    }
    \label{fig:0vbb_spectrum}
\end{figure}

The reconstructed $\hat{z}_{c}$ and $\hat{l}_e$ are applied to correct the energy spectrum event by event for both the $0\nu\beta\beta$ and $\gamma$ datasets, the performance is shown in Figure~\ref{fig:0vbb_spectrum}.
The $E_r$ is the raw energy of the event affected by the electron loss, and $E_0$ is that of the zero-attachment scenario. 
The $l_e$ of the example dataset is 1200~cm.
The difference between $E_r$ and $E_0$ represents the electron loss due to the limited electron lifetime. 
The shape of the $E_r$ spectrum is distorted and the energy peak becomes blurred for both datasets.
The $E_r$ could be corrected to $E_c$ by VGGZ0net, of which the spectrum shape is comparable with that of the $E_0$ spectrum.
In this way, by just using the background events or calibration sources, the $z_c$ reconstruction accuracy of 11~cm is enough to determine the electron lifetime and realize the energy correction.
It is beneficial to calibrate and monitor the detector performance, especially during a long-term operation. 

For different electron lifetime scenarios, the energy correction with the trained VGGZ0net is studied.
Seven datasets of $0\nu\beta\beta$ events under different gas attachment coefficients are simulated and then undergoing the above procedure. The results are shown in Table~\ref{tab:cnn_0vbb_results}.
$\sigma(\Delta z)$ is found to be independent of the electron lifetime, indicating that the model is insensitive to the carried energy of each event and only makes predictions based on the diffusion of tracks.
Both $\hat{l}_e$ and $\hat{\tau}_e$ match the defined values well, and the energy resolution is significantly improved after the correction based on VGGZ0net: 
the energy resolution at $Q_{\beta \beta}$ is improved from 10.1\% to 4.0\%~FHWM for the scenario shown in Figure~\ref{fig:0vbb_spectrum} Top.
The energy resolution of different scenarios before the correction is not given in Table~\ref{tab:cnn_0vbb_results} because the energy peak has been very blurred under a lower electron lifetime and it is difficult to fit due to the severe deformation. 
Effectively, the improved energy resolution is a more valuable index to evaluate the performance of VGGZ0net. 
As listed in Table~\ref{tab:cnn_0vbb_results}, the corrected energy resolution of the 7 datasets is comparable to that of the zero-attachment scenario.
However, the energy correction becomes more difficult to recover the energy resolution due to the fluctuations of electron loss for the events of different vertex positions when the electron lifetime is too low.

\begin{table}[htp]
    \centering
    \caption{The performance of vertex reconstruction and energy correction based on VGGZ0net in different electron lifetime scenarios. The corrected energy resolution at $Q_{\beta \beta}$ is presented.}
    \begin{tabular}{*{5}{c}}
        \toprule
        \midrule
        \makecell{$l_e$ \\ (cm)}  & \makecell{$\sigma(\Delta z)$ \\ (cm)} &  \makecell{$\hat{l}_e$ \\ (cm)}  &  \makecell{$\hat{\tau}_e$ \\ (ms)}& \makecell{ Corrected energy resolution \\ at $Q_{\beta\beta}$ (\%)~FWHM} \\
        \midrule
        Infinity & 11 & - &  - &3.3 \\
        2000   & 11&  2015 $\pm$ 55 & 10.83 $\pm$ 0.30 & 3.4 \\
        1800  & 11 &  1815 $\pm$ 53 &  9.76 $\pm$ 0.28 & 3.5 \\
        1600  & 11&  1614 $\pm$ 42 &  8.68 $\pm$ 0.23 & 3.6 \\
        1400   & 11 &  1408 $\pm$ 33  &  7.57 $\pm$ 0.18 & 3.7 \\
        1200   & 11&  1217 $\pm$ 30  & 6.54 $\pm$ 0.16 &  4.0 \\
        1000  & 11&  1008 $\pm$ 25  &  5.42 $\pm$ 0.13 & 4.2 \\
        800    & 11 &  809 $\pm$ 20  &  4.35 $\pm$ 0.11 & 4.6 \\
        \midrule
        \bottomrule
    \end{tabular}
    \label{tab:cnn_0vbb_results}
\end{table}

\section{Application and performance test in experiment}
\label{sec:Kr}

\begin{figure}[h]
		\centering
		\includegraphics[width=0.7\textwidth]{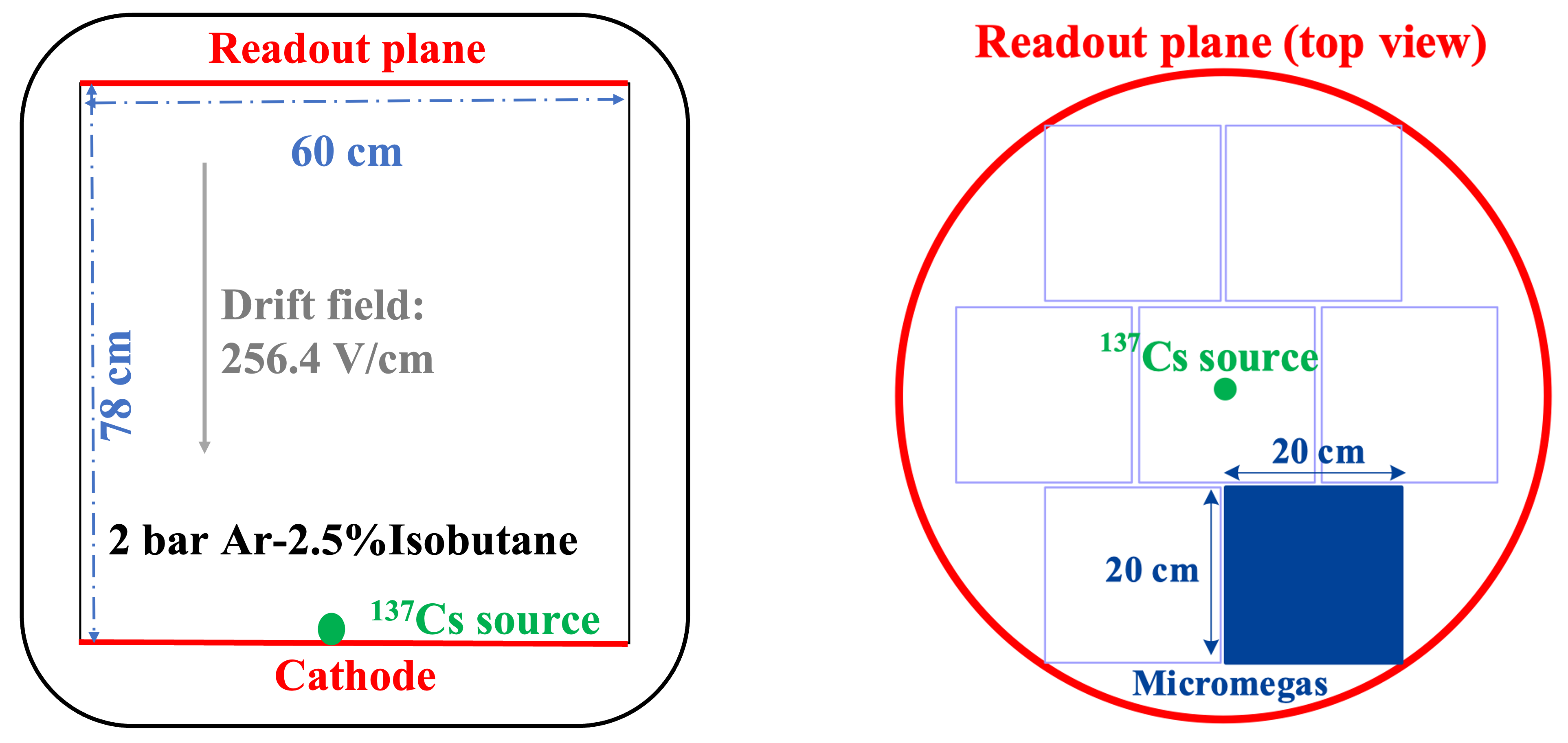}
    \caption{(Left)~Structure of the PandaX-III prototype detector~\cite{Lin:2018mpd}. The maximum drift distance of detector is 78~cm; the $^{137}$Cs source~(Green point) is placed on the cathode plane. (Right)~Structure of the readout plane. The location of the Micromegas for data taking is in blue; the relative position between the Micromegas and the projection of $^{137}$Cs source is shown.
	} 
    \label{fig:Prototype} 
\end{figure}

The PandaX-III prototype detector is built to study the performance of high-pressure xenon TPC.
A detailed description of the detector components and
subsystems is given in ref~\cite{Lin:2018mpd}.  As shown in Figure~\ref{fig:Prototype} Left, the detector vessel has an inner volume of
about 600~$liters$ and the active volume inside the TPC is about 270~$liters$, with a maximum drift distance of 78~cm.
The readout plane is composed of seven 20$\times$20~cm$^2$ Micromegas modules.
The detector has operated with different working gases and several calibration sources~\cite{Lin:2018mpd}.
For the validation of VGGZ0net, one of the seven Micromegas at the bottom right of the readout plane was selected, as shown in Figure~\ref{fig:Prototype} Right.
The detector was filled with 2~bar Ar-2.5\% isobutane mixture.
We applied a  high voltage of -20~kV on the cathode to form a drift electric field of 256.4~V/cm. Under this condition, the transverse (longitudinal) diffusion coefficient is 4.5~(2.1)$\times$10$^{-2}$~cm$^{1/2}$, 
and $v_{drift}$ is 30.3~mm/$\mu$s.
A $^{137}$Cs source was placed on the cathode and below the center Micromegas. 
At the same time, the internal calibration gaseous source $^{83m}$Kr was injected into the detector with the gas circulation through the gas handling system, generating the uniformly distributed, 41.5 keV monoenergetic events with a rate of about 1 Hz.
The strip signals of Micromegas are collected and digitized by a commercial waveform-sampling electronic system AsAd/CoBo~\cite{Pollacco:2018bsd} with a sampling rate of
5 MHz and a record length of 512 sample points, corresponding to a time bin of 0.2~$\mu$s and a time window of 102.4~$\mu$s.

\begin{figure}[h]
    \begin{subfigure}[h]{.5\textwidth}
		\centering
		\includegraphics[width=1.\textwidth]{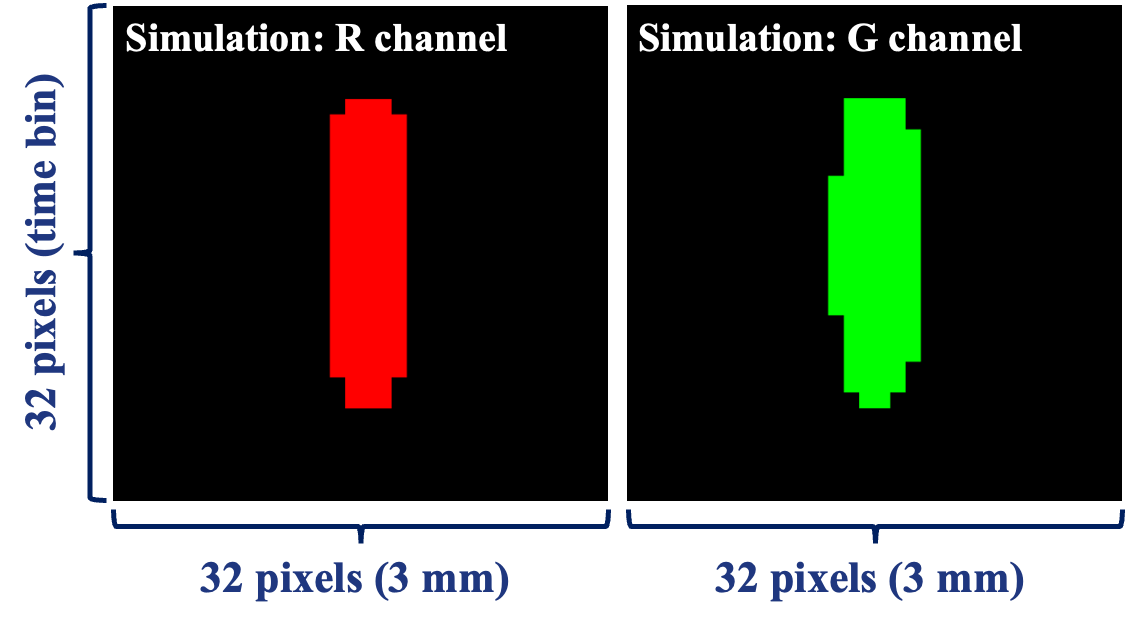}
    \end{subfigure}  
	\begin{subfigure}[h]{.5\textwidth}
		\centering
		\includegraphics[width=1.\textwidth]{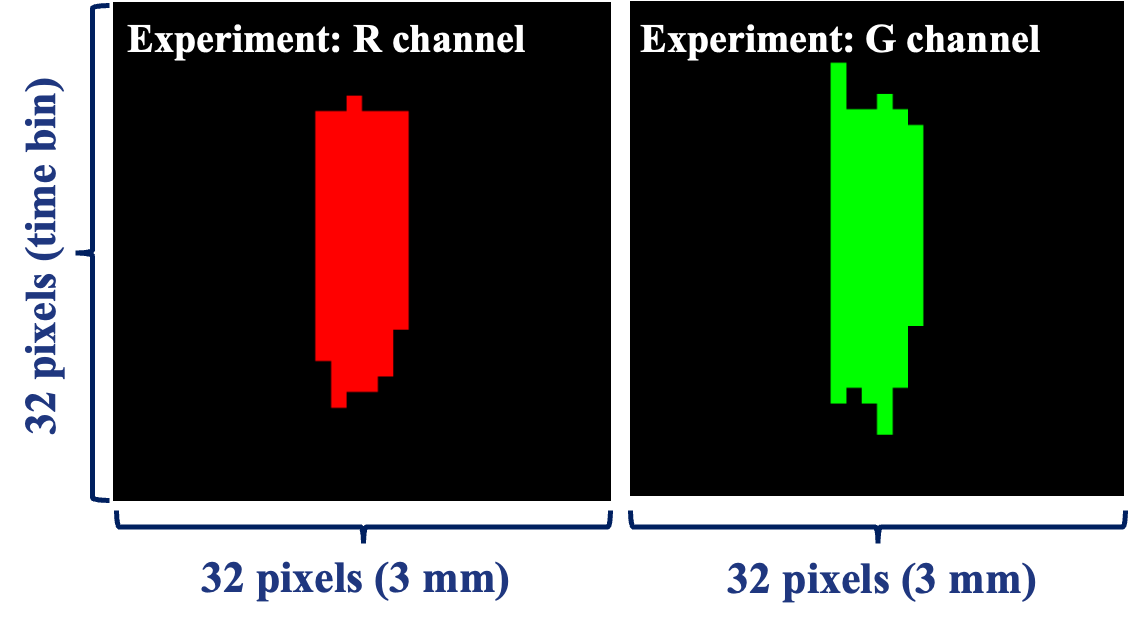}
    \end{subfigure}      
    \caption{
	Examples of RGB images of $^{83m}$Kr, the left one is from MC and the right one is from the experimental data.
	All the pixels take the maximum values and the two channels (R and G) are presented separately for better visualization.
	} 
    \label{fig:Kr_png} 
\end{figure}

Specific adjustments of the input image size and the VGGZ0net model are performed in the dozens of keV energy range.
MC datasets of $^{83m}$Kr are simulated according to the configurations of data acquisition.
The $^{83m}$Kr event tracks of MC and data are shown in Figure~\ref{fig:Kr_png}, the tracks in R and G channels are presented separately for better visualization.
The 41.5~keV events of $^{83m}$Kr triggered about 5 channels in $x$ or $y$ direction and 20 time bins in $z$ direction, leading to a cluster-like electron track.
Therefore, the size of the RGB image is reduced to 32$\times$32 to increase the proportion of pixels occupied by the short tracks. We directly reserved the $x$($y$) strips and sampling points as the pixels. 
Meanwhile, a preprocessing step is added to convert the image size back to 64$\times$64 for matching the input size of VGGZ0net.

\begin{figure}[h]
	\centering
	\includegraphics[width=.6\textwidth]{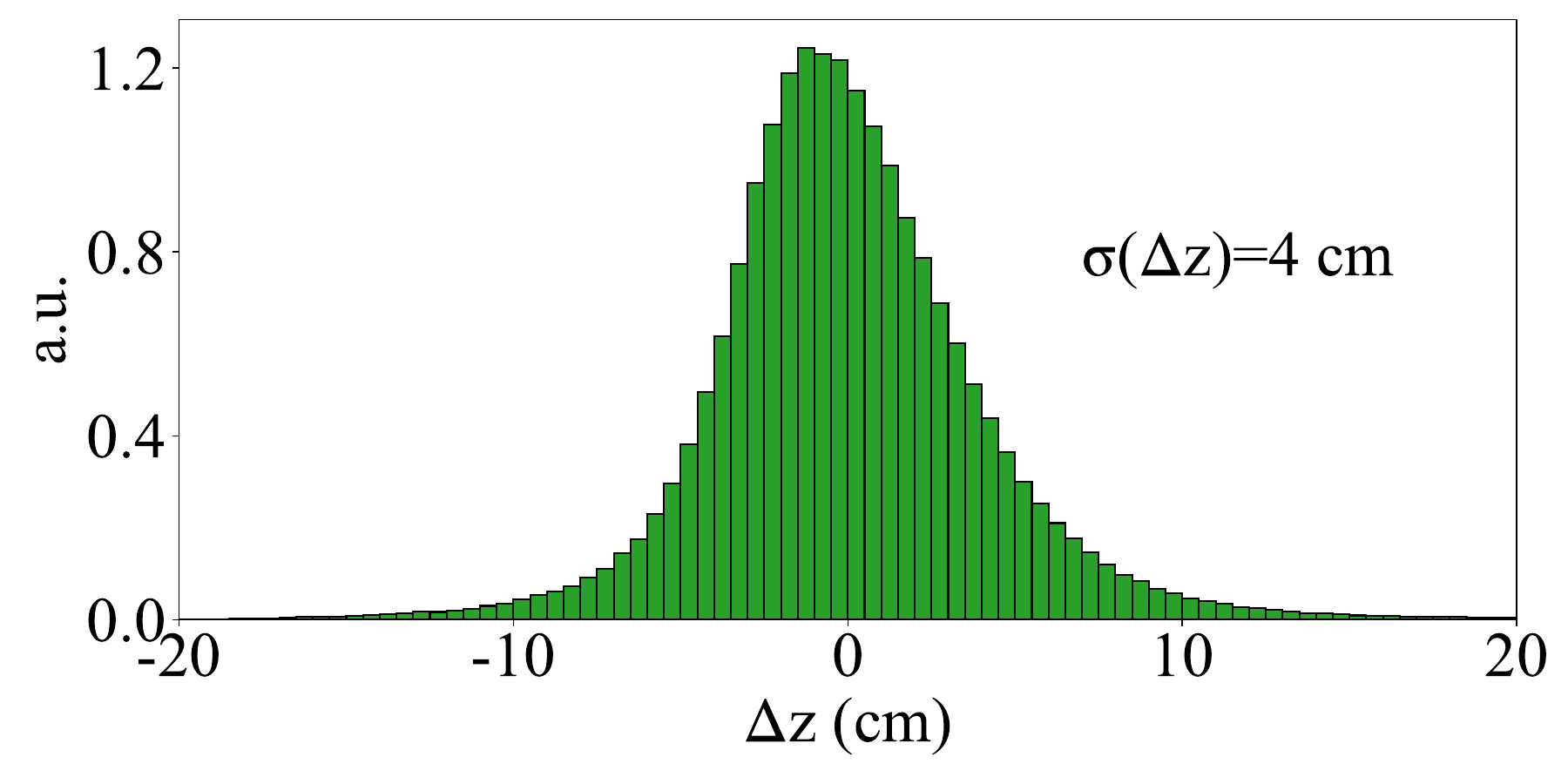} 
    \caption{The $z_c$ reconstruction precision by VGGZ0net $\sigma(\Delta z)$ is 4~cm for a simulated $^{83m}$Kr dataset.}
    \label{fig:Kr_deltaT} 
\end{figure}
The model is first trained with MC data of $^{83m}$Kr with a uniform distribution along the $z$ direction. 
The distribution of $\Delta z$ is shown in Figure~\ref{fig:Kr_deltaT} with a standard deviation $\sigma (\Delta z)$ is of 4~cm.
It is better than that of MeV-scale events presented in Section~\ref{sec:simulation}, which could be explained by the fact that the cluster-like tracks can better characterize the electron diffusion, while the features will be affected by the zigzag of long tracks in a higher energy range.

\begin{figure}[h]
		\centering
		\includegraphics[width=.8\textwidth]{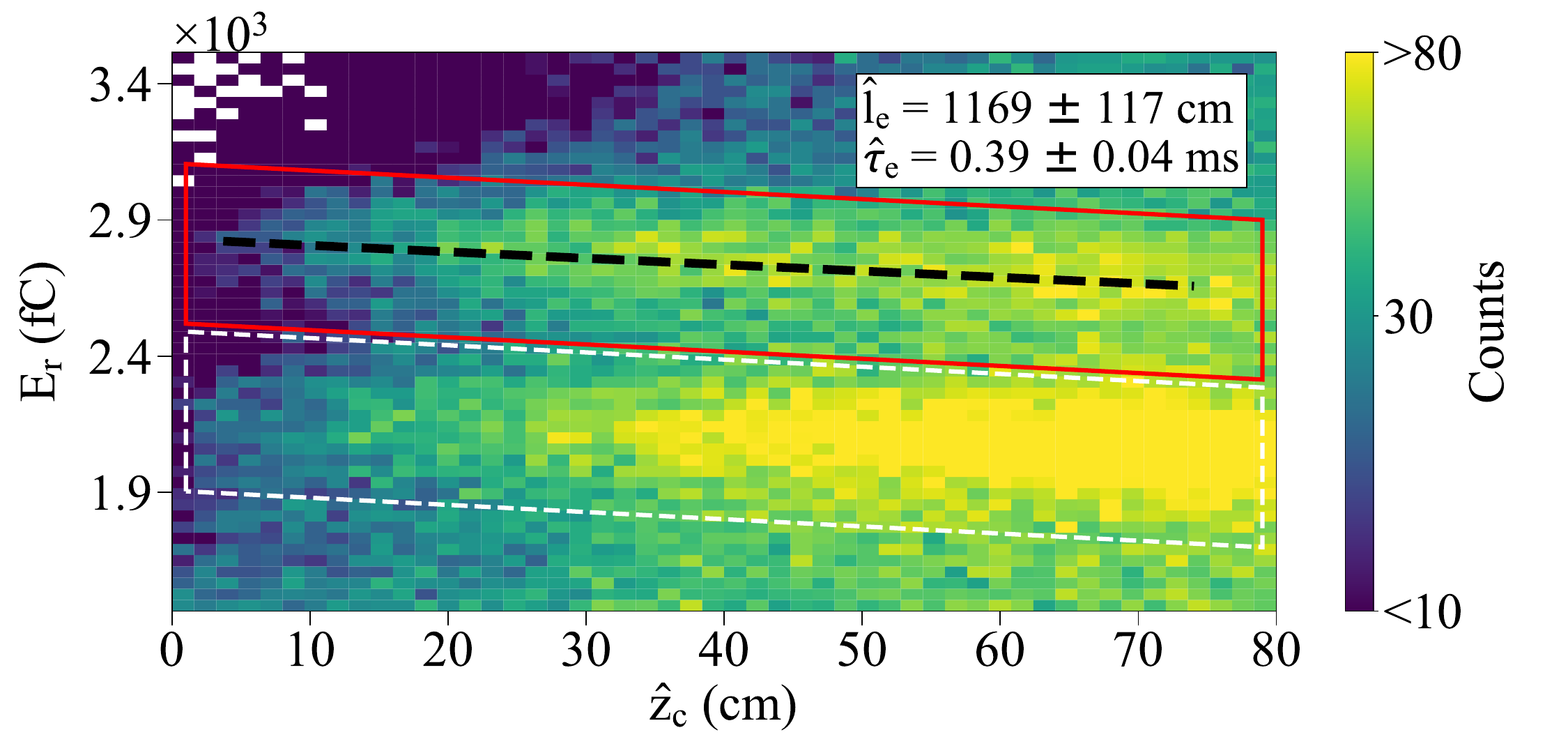}
    \caption{The distribution of detected charge $E_r$ along $\hat{z}_c$ of experimental data. 
	The $^{83m}$Kr band of 41.5~keV is shown by the red box, and the $^{137}$Cs of 32.2~keV is shown by the white dashed box.
    } 
    \label{fig:Kr_results} 
\end{figure}

The trained model is applied to the experimental data.
The distribution of the raw energy $E_r$ obtained by the detector along $\hat{z}_c$ is shown in Figure~\ref{fig:Kr_results}. 
The events within the $^{83m}$Kr band are generally distributed uniformly in the whole interval, while a large number of the events within the $^{137}$Cs band of 32.2~keV is near the cathode.
It is consistent with the expectation based on the location of the two radioactive sources.
The reconstructed $\hat{l}_e$ is 1169$\pm$117~cm from the fit to the $^{83m}$Kr band. The fitting error is generally introduced by a limited data sample. 
The equivalent $\hat{\tau}_e$ is 0.39$\pm$0.04~ms.
Thus, the trained VGGZ0net based on the MC dataset is efficient to determine $z_c$ and electron lifetime in the experimental data.
As a note, the reconstruction may have a possible bias due to the deviation between the data and MC, such as the small differences in pressure and gas ratio.  
A more quantitative optimization of the conformity can be done by tuning the detector response model in MC based on the calibration data, which is left for future work.

\section{Summary}
\label{sec:conclusion}

In this paper, we report the event vertex reconstruction using CNN in the high-pressure TPC of the PandaX-III experiment. 
The lack of the event vertex along the drift direction will directly affect the energy resolution of the detector and the background rejection,
which are the key parameters for searching $0\nu\beta\beta$.
As the scintillation signal cannot be measured in the xenon-TMA gas mixture, the diffusion information carried by the tracks is an efficient and unique signature to reconstruct the vertex. 
VGGZ0net is built to predict the vertex of the events distributed continuously between the cathode and the readout plane within the TPC. 
MC simulation datasets of $0 \nu \beta \beta$ events are prepared to train the CNN model and tune its hyperparameters. 
The reconstruction accuracy of VGGZ0net can reach a standard deviation of about 11~cm.
On this basis, the electron lifetime can be derived and the energy spectrum can be corrected for both the $0 \nu \beta \beta$ and $\gamma$ datasets.
VGGZ0net is proven to be efficient and the improvement of energy resolution is significant, for example, from 10.1\% to 4.0\%~FWHM at $Q_{\beta\beta}$ with the $l_e$ of 1200~$cm$. This method has been verified by the experimental data of the PandaX-III prototype detector using the $^{83m}$Kr calibration run. 
In addition, once the vertex position is known, the radioactive background from the detector construction materials near the readout plane and the cathode can be identified and rejected by reconstructing its location.
This method will be further verified and optimized when the PandaX-III experiment takes data in the future.

\acknowledgments

This work is supported by the grants from National Natural Sciences Foundation of China
~(No.11775142, and No.11905127). 
We thank the support from the Key Laboratory for Particle Physics, Astrophysics and Cosmology, Ministry of Education.
This work is supported in part by the Chinese Academy of Sciences Center for Excellence in Particle Physics~(CCEPP).

\bibliographystyle{JHEP}
\bibliography{main}

\end{document}